# Tunable reactivity of supported single metal atoms by impurity engineering of the MgO(001) support

Igor A. Pašti[1,2]*, Börje Johansson[2,3], Natalia V. Skorodumova[2,3]

[1]*University of Belgrade – Faculty of Physical Chemistry, Belgrade, Serbia*

[2]*Department of Materials Science and Engineering, School of Industrial Engineering and Management, KTH - Royal Institute of Technology, Stockholm, Sweden*

[3]*Department of Physics and Astronomy, Uppsala University, Uppsala, Sweden*

* **Corresponding author, e-mail: igor@ffh.bg.ac.rs**

**Abstract**

Development of novel materials may often require a rational use of high price components, like noble metals, in combination with the possibility to tune their properties in a desirable way. Here we present a theoretical DFT study of Au and Pd single atoms supported by doped MgO(001). By introducing B, C and N impurities into the MgO(001) surface, the interaction between the surface and the supported metal adatoms can be adjusted. Impurity atoms act as strong binding sites for Au and Pd adatoms and can help to produce highly dispersed metal particles. The reactivity of metal atoms supported by doped MgO(001), as probed by CO, is altered compared to their counterparts on pristine MgO(001). We find that Pd atoms on doped MgO(001) are less reactive than on perfect MgO(001). In contrast, Au adatoms bind CO much stronger when placed on doped MgO(001). In the case of Au on N-doped MgO(001) we find that charge redistribution between the metal atom and impurity takes place even when not in direct contact, which enhances the interaction of Au with CO. The presented results suggest possible ways for optimizing the reactivity of oxide supported metal catalysts through impurity engineering.

**Keywords:** magnesium oxide; surface doping; surface reactivity; CO adsorption; supported metal atoms

## 1. Introduction

Nowadays, catalysis impacts the areas of energy, pollution control, and chemical production, with more than 90% of commercial processes employing catalysts to accelerate the rate of chemical transformation into desired products. The selection of a catalyst is governed by a balance between the catalytic performance and price. Hence, the reduction of the content of expensive components, like noble metals, is one of top priorities. Supported single metal atoms and small metal clusters have been used as academic play models for the understanding of cluster growth, nucleation, mobility and catalytic activity.[1] However, many recent reports suggest tremendous development of catalysts based on small metal clusters and low coordinated atoms.[2] While it is known that low coordinated atoms can often provide enhanced catalytic activities[3] single atoms are even more attractive when each atom bears catalytic functions.[4-7] One of the main challenges in single atom catalysis is a proper anchoring of single atoms to the support.[8,9] Therefore, tuning the properties of the support can be suitable strategy to modify the activity and stability of single atom catalysts.[10]

Oxide materials show versatile properties and they are often used as catalyst supports. Among them, magnesium oxide, MgO, is of particular importance.[11-14] Its most stable surface, MgO(001), is non-polar, easy to prepare and does not undergo considerable structural relaxation.[15,16] Due to the desirable properties of MgO, it is widely used in surface science and catalysis. This has resulted in a large body of theoretical and experimental works focusing on the interactions of metals and small clusters with MgO(001).[17-23] The properties of supported metal atoms and clusters are shown to be affected by properties of the substrate.[24] In particular, the presence of defects[25] and deposition of atoms onto thin oxide films supported by metallic substrates[26] alter the growth and properties of metal atoms and metal clusters, affecting their reactivity.[27] Among other properties, particular attention has been paid to the impact of the particle charge state on the chemical properties.[12, 28-30] It is suggested that an extra charge can boost bond cleavage in molecular adsorbates[31] and alter chemisorption properties of supported metals.[32] For these reasons, low coordinated atoms on oxide substrates are interesting as their charge can be controlled either by substrate doping[33-35] or by using a support in the form of a thin oxide film deposited on a suitable metallic substrate.[36-38] An alteration of the properties of oxide supports can also be accomplished by doping them with non-metals. For example, the appearance of the so-called $d^0$ magnetism in N- and C-doped MgO has received significant attention.[39-41] In addition, we have recently shown that doping of MgO(001) with B, C and N alters the surface reactivity as well.[42] The origin of the altered reactivity is due to unpaired electrons localized at dopant atoms, presenting surface embedded radical species.[42]

In the present work we extend our previous study on B-, C- and N-doped MgO(001) and consider these modified surfaces as possible support for Pd and Au adatoms. The selection of adatoms is based on the large amount of work done so far,[22,24,43] providing clear-cut reference points. Considering the importance of the CO oxidation[44-46] and the significant role of small metal clusters in this catalytic process,[12,47] the reactivity of supported Au and Pd atoms are tested using the CO molecule as a probe.

## 2. Computational details

The calculations were based on DFT within the generalized gradient approximation, using Perdew–Burke–Ernzerhof exchange correlation functional.[48] The calculations were performed by means of the Quantum ESPRESSO *ab initio* package.[49] For Mg, O, B, C and N, the $n$s and $n$p states were treated as the valence states ($n$ = 3 for Mg and $n$ = 2 for O, B, C and N). For the case of Pd, the s- and d-states were included into the valence band, while for Au also the p-states were taken into account. The basis set was expanded by the plane waves with a maximum kinetic energy of 28 Ry. The charge density cutoff was 16 times higher. Spin polarization was taken into account for all the investigated systems. The calculated equilibrium lattice constant of MgO was 4.22 Å, in good agreement with the experimental value (4.21 Å[50]).

The MgO (001) surface was modeled by a 2x2 four layer thick slab. One dopant atom (B, C or N) *per* simulation cell was introduced. As previously shown, these dopants prefer to be located in the surface layer and only such configurations of doped surface were considered.[42] The concentration of dopants in the simulation cell is 1.56 at.%, which is below the concentrations usually reported experimentally for N-doped MgO (up to 6 at.% according to Ref.([39])). The lateral dimensions of the supercell were constrained to those of pristine MgO(001). This resulted in a certain surface stress due to the introduction of impurities. However, we chose to compare surface reactivity by keeping the lattice constant to that of MgO. The first irreducible Brillouin zone was integrated using a 4×4×1 Monkhorst-Pack grid.[51] A Gaussian smearing procedure, with a broadening of 0.007 Ry, was applied. All the atoms in the simulation cell were allowed to relax, except for the first bottom layer, which was fixed during geometry optimization. The surface slabs were separated by a vacuum region of 20 Å. The dipole correction was applied to prevent coupling of periodic images along the *z*-direction.[52]

In order to quantify the strength of the interaction between the support and metal adatom the binding energy ($E_b(M)$) is used:

$$E_b(M) = E_{M@X\text{-MgO(001)}} - E_{X\text{-MgO(001)}} - E_M \qquad (1)$$

where $E_{M@X\text{-MgO(001)}}$, $E_{X\text{-MgO(001)}}$ and $E_M$ stand for the total energies of M (Pd or Au) adsorbed on a given substrate, clean X-doped MgO(001) (X = B, C, N or O, the last one refers to pristine MgO(001)) and the isolated atom M, respectively. Binding of Pd and Au atoms to the support was probed at the impurity sites and neighboring anionic O sites (Fig. 1, O 1N, O 2N and O 3N sites). Additional test calculations were performed using a larger, 3×3 cell, which was constructed as a three layer slab, in order to check for the interactions between periodic images and the lateral distances between adatoms and impurities.

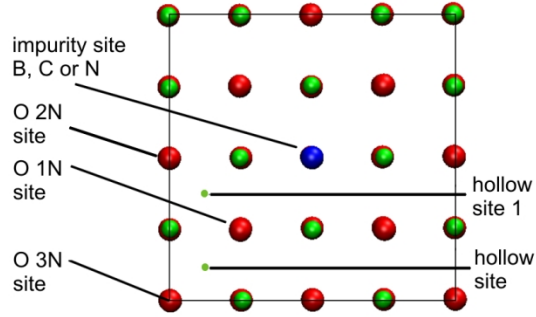

**Figure 1**. Notation of metal adsorption sites investigated in the present work

The reactivity of supported metal atoms was probed using the CO molecule. The chemisorption of CO is quantified as the adsorption energy ($E_{ads}(CO)$):

$$E_{ads}(CO) = E_{CO@M@X-MgO(001)} - E_{M@X-MgO(001)} - E_{CO} \quad (2)$$

where $E_{CO@M@X-MgO(001)}$, and $E_{CO}$ stand for the total energy of CO adsorbed on supported metal atom and the total energy of isolated CO molecule, respectively. The charge transfer was analyzed using the Bader algorithm[53] on the charge density grid by Henkelman *et al.*[54] For graphical presentation the VMD code was used.[55]

### 3. Results and discussion

3.1. Doped MgO as the support for metal adatoms

Both Pd and Au interact rather weakly with pristine MgO(001) terrace and prefer the anionic O site for adsorption (Table 1). The obtained results indicate that in both cases charge is transferred to the adatoms. Pd receives 0.25e, while 0.30e is transferred to Au. The obtained results agree very well with the ones reported previously by various authors, as summarized in Table 1. In addition, both Au and Pd are known to be quite mobile on the MgO(001) surface.[22,56] The saddle point for diffusion on MgO(001) is the hollow site (Fig. 1)[56] and we estimate the diffusion barriers of Pd and Au on pristine MgO(001) to be 0.37 eV and 0.27 eV, respectively.

The barriers were estimated as a the difference between the calculated binding energies of Pd(Au) at the O site and the hollow site. Our results agree well with previous reports: 0.34 eV for the Pd adatom[22] and 0.22 eV for Au.[56] The estimated barrier for Au diffusion also falls into the range predicted experimentally (0.2-0.3 eV).[56]

**Table 1.** Adsorption of Pd and Au on pristine MgO(001). Metal binding energies, the distances between Pd/Au and O site, charge of metal adatom and ground state magnetizations are provided. Comparison with the literature data is given. Results obtained in present work are underlined.

| Adsorbed metal | adsorption site | $E_b(M)$ / eV | $d(Pd–O)$ / Å | $q(M)^*$ / \|e\| | $M$ / $\mu_B$ |
|---|---|---|---|---|---|
| Pd | O top | <u>−1.35</u> | <u>2.10</u> | <u>−0.25</u> | <u>0</u> |
|  |  | −1.4[a] |  |  |  |
|  |  | −1.02[b] | 2.08[b] |  | 0[b] |
|  |  | −1.36[c] | 2.09[c] |  | 0[c] |
|  |  | −1.35[d] | 2.10[d] | −0.24[d] | 0[d] |
| Au | O top | <u>−0.91</u> | <u>2.25</u> | <u>−0.30</u> | <u>1</u> |
|  |  | −0.74[b] | 2.24 |  | 1[b] |
|  |  | −0.88[c] | 2.29[c] |  | 1[c] |
|  |  | −0.89[e] | 2.32[e] | −0.30[e] | 1[e] |
|  |  | −0.78[f] | 2.33[f] | −0.31[f] |  |
|  |  | −1.01[g] | 2.24[g] |  |  |
|  |  | −0.91[h] |  |  |  |

*negative values indicate an excess electron charge on adatoms.
[a]Ref. [22]; [b]PBEN results, cluster approach [57]; [c]Ref. [58], [d]Ref. [42]; [e]Ref. [11], [f]BSSE-corrected energies [59]; [g]Ref. [60]; [h]Ref. [56]

Next, we turn to metal binding on doped surfaces. As previously shown, the investigated impurities prefer to be located in the surface layer, rather than beneath the surface.[42] Due to the unpaired electrons on dopant atoms the impurity sites were found to bind Pd and Au strongly, with $E_b$(Au) reaching −3.72 eV in the case of B-doped MgO(001) (Table 2). Metal binding becomes weaker as going from B to O site (pristine MgO(001)), following the decrease in the (formal) number of unpaired electrons at the adsorption site. Doping of MgO(001) with B, C and N gives rise to the magnetic moments of 3, 2 and 1 $\mu_B$, respectively, which are predominantly localized at the dopants.[42] Bonding of Pd and Au on B-doped MgO(001) is preferred in a tilted configuration, where metal adatoms are tilted towards $Mg^{2+}$ centers. Such an orientation can be understood on the basis of the charge state of Pd and Au adatoms, as in the case of B-doped surface a significant charge is transferred to metal adatoms (Table 2).

Table 2. Adsorption of Pd and Au on doped MgO(001) – ground states for metal adsorption at impurity sites. Metal binding energies, the distances between Pd/Au and the impurity site, charge of metal adatom and ground state magnetizations are provided.

| Adsorbed metal | Substrate | adsorption site | $E_b(M)$ / eV | $d(M–X)$ / Å | $q(M)$ / \|e\| | $M$ / $\mu_B$ |
|---|---|---|---|---|---|---|
| Pd | B-MgO(001) | B tilted | −3.64 | 1.88 | −0.79 | 1 |
|    | C-MgO(001) | C top    | −3.64 | 1.90 | −0.19 | 2 |
|    | N-MgO(001) | N top    | −2.69 | 1.96 | −0.08 | 1 |
| Au | B-MgO(001) | B tilted | −3.72 | 1.97 | −1.46 | 2 |
|    | C-MgO(001) | C top    | −3.37 | 1.94 | −0.16 | 1 |
|    | N-MgO(001) | N top    | −2.89 | 1.97 | +0.01 | 0 |

Compared to adsorption on pristine MgO(001), the binding of Au and Pd to C- and N-doped MgO(001) results in less charge transferred to the adatoms. However, the magnetic properties (Table 2) and the electronic structures (Figs. 2 and 3) of metal adatoms are also affected by the presence of surface impurities. The ground state of a Pd atom is a singlet and its binding to C- and N-doped surface does not change the magnetic moment of system. However, in the case of B-doped surface, the coupling between Pd states and B states reduces the magnetic moment of the system down to 1 $\mu_B$. In the case of Au, one unpaired electron of Au combines with the unpaired electrons of the impurity and reduces the magnetic moment of each system by 1 $\mu_B$. While the magnetic moments of doped MgO(001) surfaces are localized at dopant atoms, in the case of adsorbed Pd and Au the magnetic moments are also found at the adatoms and surface atoms around them (Figs. 2 and 3).

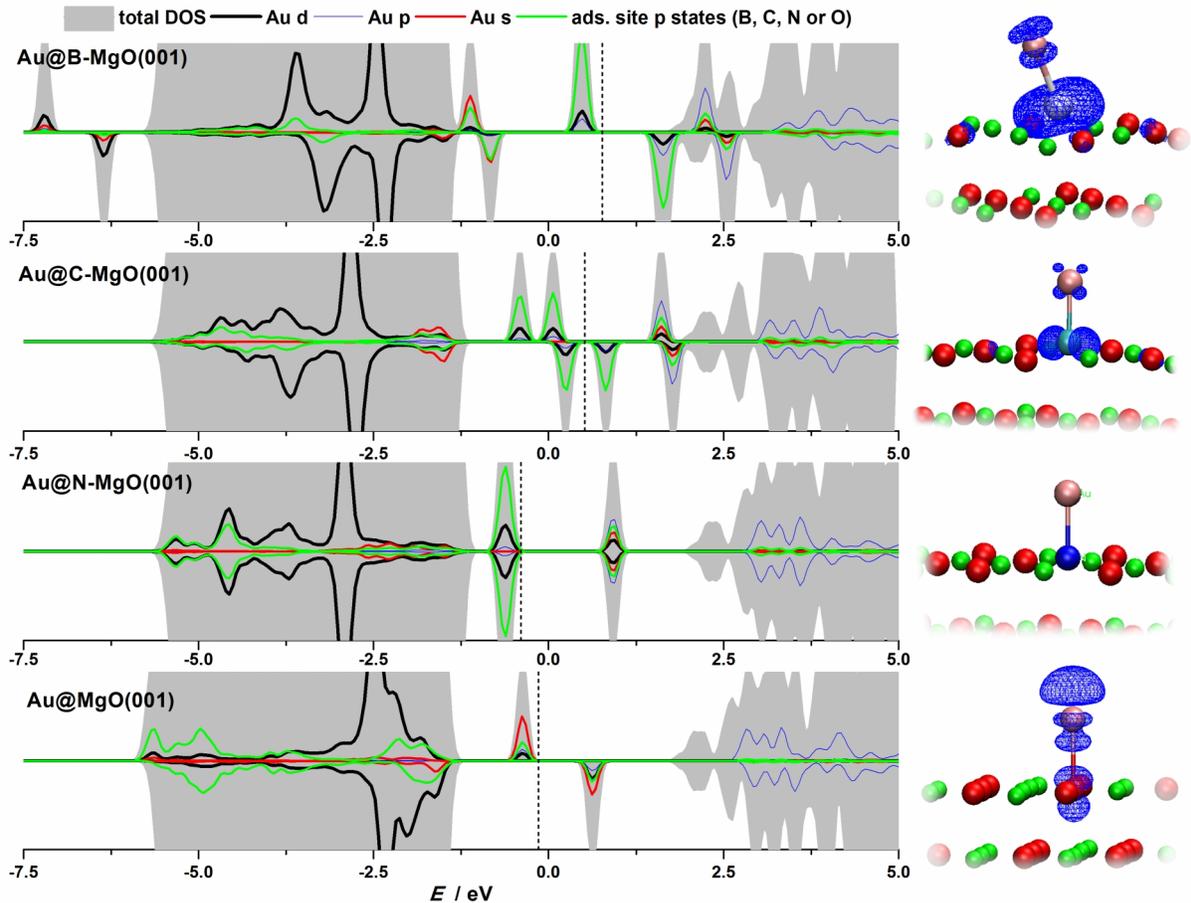

**Figure 2**. Projected densities of states for Au adatoms on doped and pristine MgO(001). Insets give the optimized structures with 3D magnetization isosurfaces (given as $\rho_\uparrow - \rho_\downarrow$). The vertical dashed lines indicate the highest occupied states.

Having analyzed the electronic structure of adatoms on doped MgO(001), we see that there are significant changes in the projected states as compared to those of metals adsorbed on pristine MgO(001) (Fig. 2 and 3). The states of adatoms (s- and d-states) couple strongly to the states of surface atoms to which they are bound (impurity or O center). In the case of pristine MgO(001) this gives rise to the states located in the band gap. However, in the case of doped MgO(001) there are states which are already present in the band gap[42] and adatoms states interact with them strongly. A strong interaction between the metal and impurity states results in splitting of the Au d-states, found in approx. 3 eV wide energy window (Fig. 2). A similar trend is seen for the case of Pd (Fig. 3). We expect that such significant alterations of the electronic structure should have a considerable impact on the reactivity of these adatoms.

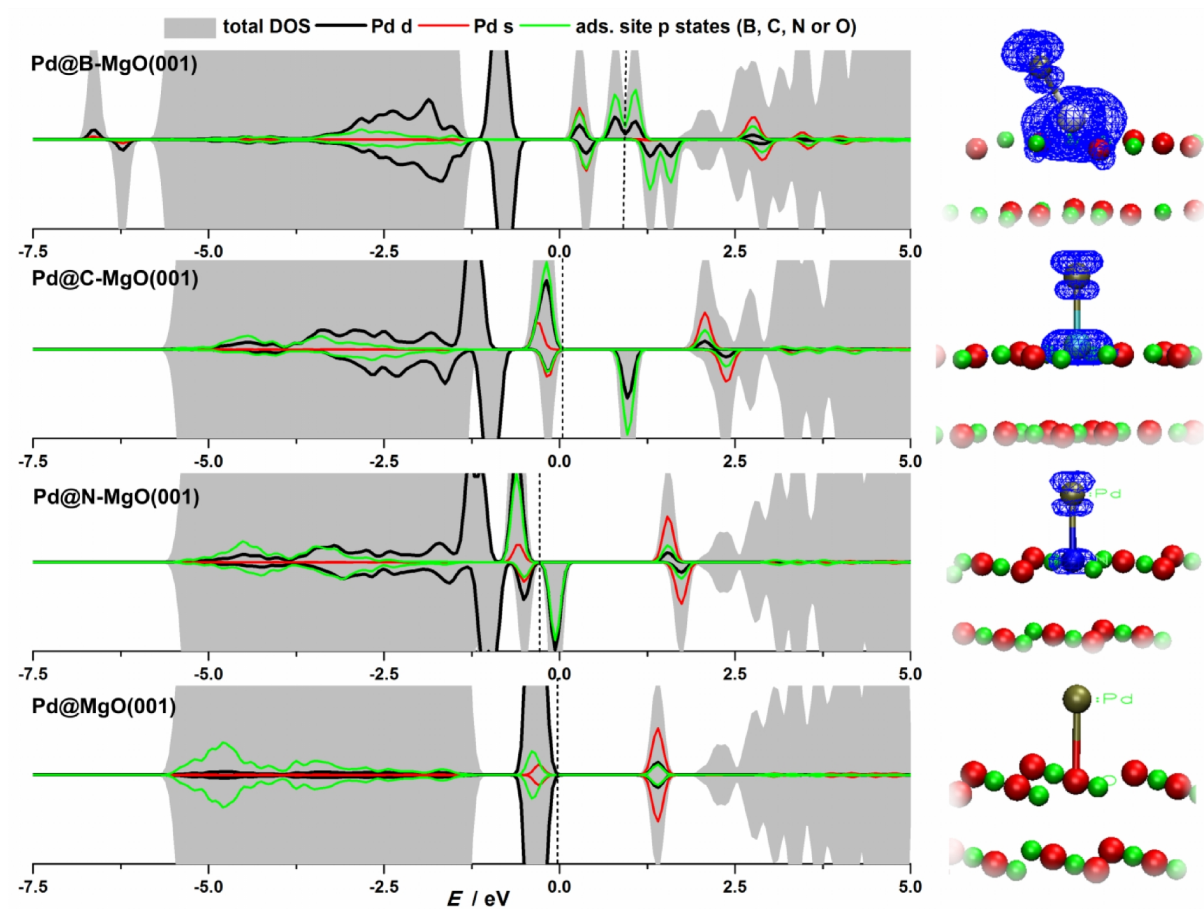

**Figure 3**. Projected densities of states for Pd adatoms on doped and pristine MgO(001). Insets give the optimized structures with 3D magnetization isosurfaces (given as $\rho_\uparrow - \rho_\downarrow$). The vertical dashed lines indicate the highest occupied states.

Besides the adsorption on impurity sites, we have also investigated the interaction of Au and Pd with the O centers of doped MgO(001). We found that first neighbor O sites (1N sites, see Fig. 1 for notations) are not stable adsorption sites for all the adatoms, as they relax towards impurity sites to the configurations described in Figs. 2 and 3 and Table 2. In the case of gold we tested adsorption at 2N ($2^{nd}$ neighbor) and 3N ($3^{rd}$ neighbor) sites (Fig. 1) and found that the corresponding binding energies are rather close to that on pristine MgO(001) (Table 3), while there are no lateral forces acting on Au adatoms.

Table 3. Adsorption of Pd and Au at doped MgO(001) – adsorption at O 3N sites. For the case of O 1N sites metal atom spontaneously relax to the impurity site. The data for the adsorption on pristine MgO(001) are also included for easier comparison.

| Adsorbed metal | Ads. site | Substrate | $E_b$(M) / eV | $d$(M–O) / Å | $q$(M) / \|e\| | $M$ / $\mu_B$ |
|---|---|---|---|---|---|---|
| Au | O 2N | B-MgO(001) | −0.84 | 2.58 | −0.46 | 2 |
|  |  | C-MgO(001) | −0.80 | 2.29 | −0.31 | 1 |
|  |  | N-MgO(001) | −0.88 | 2.26 | −0.30 | 0 |
| Au | O 3N | B-MgO(001) | −0.81 | 2.31 | −0.31 | 2 |
|  |  | C-MgO(001) | −0.85 | 2.27 | −0.30 | 1 |
|  |  | N-MgO(001) | −0.89 | 2.26 | −0.30 | 2 |
| Pd | O 3N | B-MgO(001) | −1.28 | 2.11 | −0.23 | 3 |
|  |  | C-MgO(001) | −1.30 | 2.11 | −0.24 | 2 |
|  |  | N-MgO(001) | −1.32 | 2.11 | −0.25 | 1 |
| Au | O | MgO(001) | −0.91 | 2.25 | −0.30 | 1 |
| Pd | O | MgO(001) | −1.35 | 2.10 | −0.25 | 0 |

We also performed test calculations using a 3×3 cell of pristine and B-doped MgO(001). In that simulation cell Au binding at the anionic O sites up to 5N (following notation given in Fig. 1) was analyzed. We observed slight variations in the Au binding energies at O sites (−0.85 to −0.89 eV) compared to that obtained for Au at pristine MgO(001) in the same simulations cell (−0.90 eV). We assume that the alteration of the metal binding energy on doped MgO(001) compared to pristine MgO(001) is due to the lateral stress induced by dopants and it is most pronounced for the case of doping with boron and it decreases when going to nitrogen. Namely, the incorporation of impurities into MgO results in the deformation of the lattice, which is most pronounced in the case of B-doping. This can be unambiguously attributed to a large ionic radius of B compared to those of C and N, as discussed in Ref. [42]. In the case of Pd adsorption we also see that the adsorption at the O 3N site is relatively weakly affected by the impurity (Table 3). The same holds for the charge states of metal adatoms.

The magnetic moments obtained for Pd@X-MgO(001) are similar to those of doped surfaces. In the case of Au@X-MgO(001) the situation is a bit more complicated (Fig. 4).

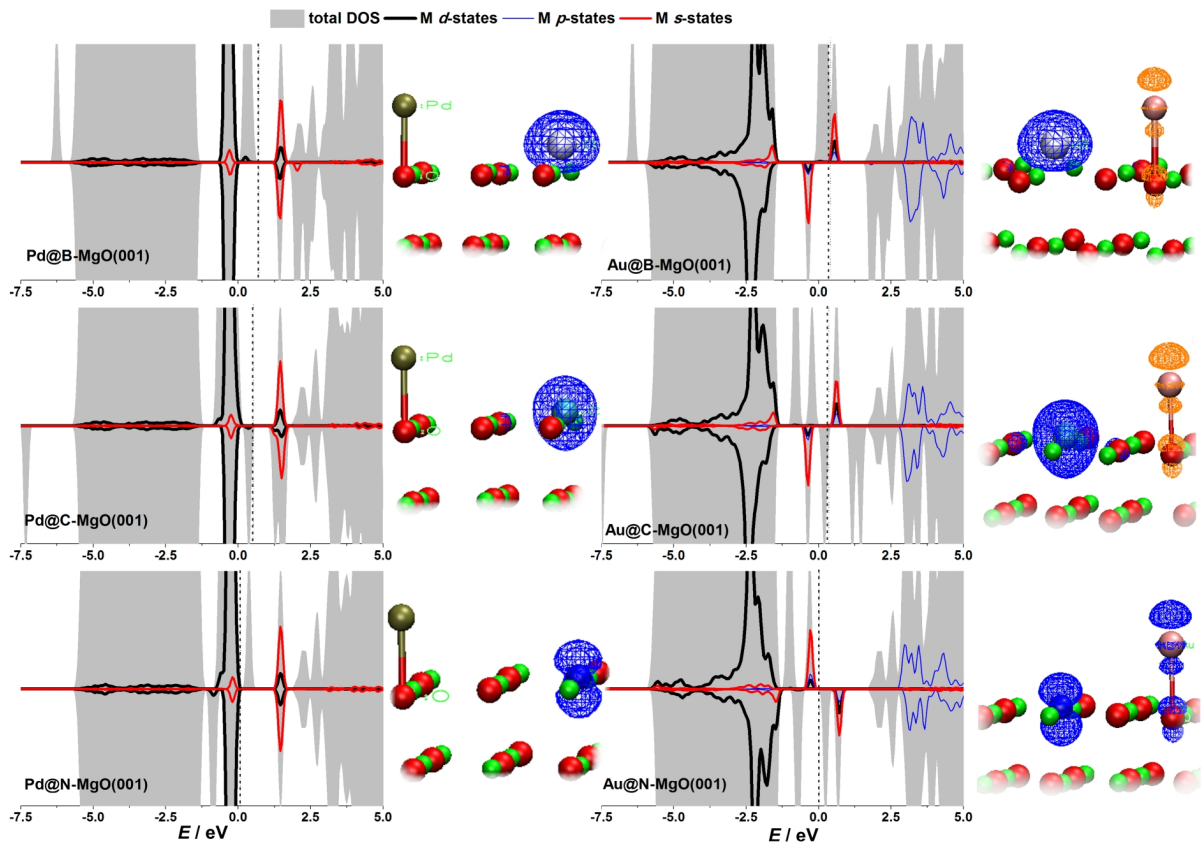

**Figure 4**. Projected densities of states for adatoms on doped and pristine MgO(001) at O 3N sites. Insets give the optimized structures with 3D magnetization isosurfaces (given as $\rho_\uparrow - \rho_\downarrow$). Positive isosurfaces are blue, negative are given in orange. The vertical dashed lines indicate the highest occupied states.

When Au is adsorbed at the O 2N site the antiferromagnetic spin arrangement between the Au adatom and the impurity reduces the total magnetization by 1 $\mu_B$. This is also the case for Au adsorbed at the O 3N site of B- and C-doped MgO(001). In contrast, for the case of Au adsorbed at the O 3N site of N-doped MgO(001) the Au magnetic moment is parallel to that of N and the total magnetization of the system is 2 $\mu_B$. The antiferromagnetic arrangement is less stable then the ferromagnetic one by 0.37 eV. The electronic structures of metal adatoms at the O 3N site (Fig. 4), is rather similar to that on pristine MgO(001) (Figs. 2 and 3). In all the cases the *d*-states of metal atoms are practically completely filled. In the case of Pd, the s-states are mainly empty, while in the case of Au ($n$s$^1$($n$−1)d$^{10}$ configuration) the s-states split into two components, one of which being filled. These states arise due to the interaction of Au with the O center and are located in the band gap (Figs. 2 and 3).

A rather strong binding of Au and Pd at the impurity sites combined with rather weak interactions with pristine MgO(001) terrace sites (starting from O 2N sites) and low diffusion barriers of adatoms suggest that Au and Pd can be effectively trapped at the impurity sites on MgO(001). To reinforce this conclusion we estimated the mobility of Pd/Au on the doped MgO(001) surfaces considering relative differences in metal binding energies at O sites and hollow sites (Fig. 1). In all cases the barriers are somewhat reduced compared to pristine MgO(001) terrace. This is especially pronounced for Au on B- and C-doped MgO(001), where we found barriers of about 0.01 eV. In addition, a comparison between Au/Pd binding energies on doped MgO(001) and the bulk cohesive energies can be made. In contrast to the case of pristine MgO(001), the binding energies at impurity sites are much closer to the cohesive energies of the considered metals (cohesive energies amount 3.89 eV atom$^{-1}$ for Pd and 3.81 eV atom$^{-1}$ for Au[61]). Hence, by combining the high mobility on MgO(001) with strong bonding at the impurity sites, we conclude that the impurity sites can serve as strong anchoring sites for individual Au and Pd atoms. The atoms trapped at the impurity sites would require significant energy (more than 1 eV in the case of Pd and close to 2 eV for Au) to leave such a site. As the binding and cohesive energies are similar, the introduction of impurities into the surface could prevent metal aggregation and result in a high catalyst dispersion.

Before proceeding further it is important to briefly discuss the possibility of experimental preparation of doped MgO surfaces and their use as catalyst supports. According to Ref. ($^{39}$) a relatively high concentration of N (up to at.%) was achieved by MgO deposition on Mo substrate. Due to the catalytic action of Mo for $N_2$ dissociation, atomic N was effectively incorporated into the MgO films during the film growth. However, scanning tunneling spectroscopy experiments have identified some features, which cannot be ascribed to single N impurities. Hence, there is a certain possibility that some complex structures involving impurities might be formed. However, we expect that this can be avoided for low concentrations of impurities, which would reduce the probability of impurity clustering. Moreover, the embedding of the studied impurities into the oxygen vacancies of MgO is highly exothermic[42], suggesting that a high energy input would be required for such an impurity to leave its lattice site and recombine with another impurity. Additionally, in order to achieve high metal (Pd, Au) dispersion, ideally a monoatomic one, the surface concentration of supported metal atoms should also be kept low and comparable to the surface concentration of non-metal (B, C, N) impurities.

3.2. Reactivity of Au and Pd supported by doped MgO

The reactivity of supported metal atoms was probed using CO. In the case of Pd@MgO(001) the obtained results are in accordance with our previous work[62] and the report by Del Vitto *et al*.[63] For the case of Au@MgO(001), CO is adsorbed relatively weakly on the Au adatom, with adsorption energies, C–O bond lengths and calculated charge transfer (Table 4) which agree with previous reports.[11,64] By comparing Pd and Au on pristine MgO(001) one can see that Pd binds CO much stronger (Table 4). Once adsorbed, CO gets negatively charged due to electron backdonation to the 2π* orbital of CO,[65] while the M-CO complex is only weakly negatively charged. It results in weakening of the C–O bond, which is also seen in its elongation (Table 4; the calculated C–O bond length in an isolated CO molecule is 1.14 Å). We note that the CO adsorption on pristine MgO(001) takes place at the Mg site with binding energy of only −0.16 eV.[42]

The interaction of CO with Pd@X-MgO(001) is much weaker compared to the case of Pd@MgO(001). For the case of B-doped MgO(001) the CO adsorption energy is reduced to only −1.09 eV. In contrast, once the Au adatom is attached to the impurity it becomes much more reactive than on pristine MgO(001). For N-doped MgO(001) the CO adsorption energy reaches −2.30 eV (Table 4). When considering the series of dopants, the CO adsorption is always the weakest on the metal adatoms attached to B and the strongest when attached to N. Interestingly, this order is opposite to the one observed for the strength of the interaction with the impurity site – Au and Pd bind strongest to B and weakest to N. We see that the charge backdonation from Pd(Au) to the CO states is still present for doped systems but to a somewhat smaller extent compared to that on pristine MgO(001) (Table 4).

Table 4. Characterization of CO adsorption on Pd and Au atoms supported by pristine MgO(001) and doped MgO(001) surfaces.

| | surface | $E_{ads}$(CO) / eV | $d$(M–X)* / Å | $d$(M–C) / Å | $d$(C–O) / Å | $q$(M) / \|e\| | $q$(CO) / \|e\| | $M$ / $\mu_B$ |
|---|---|---|---|---|---|---|---|---|
| **Pd** | MgO | −2.44 | 2.10 | 1.87 | 1.16 | +0.05 | −0.23 | 0 |
| **Au** | MgO | −0.97 | 2.06 | 1.88 | 1.18 | +0.12 | −0.29 | 1 |
| **Pd** | | | | | | | | |
| impurity site | B-MgO | −1.09 | 1.96 | 1.98 | 1.16 | −0.26 | −0.23 | 1 |
| | C-MgO | −1.29 | 1.97 | 1.98 | 1.16 | +0.10 | −0.21 | 2 |
| | N-MgO | −1.83 | 2.00 | 1.93 | 1.16 | +0.04 | −0.18 | 1 |
| O 3N site | B-MgO | −2.36 | 2.11 | 1.88 | 1.16 | +0.03 | −0.20 | 3 |
| | C-MgO | −2.41 | 2.11 | 1.88 | 1.16 | +0.02 | −0.22 | 2 |
| | N-MgO | −2.43 | 2.11 | 1.88 | 1.16 | +0.04 | −0.23 | 1 |
| **Au** | | | | | | | | |
| impurity site | B-MgO | −1.00 | 2.01 | 1.99 | 1.16 | −0.25 | −0.15 | 2 |
| | C-MgO | −1.67 | 1.95 | 1.93 | 1.16 | +0.18 | −0.20 | 1 |
| | N-MgO | −2.30 | 1.94 | 1.88 | 1.16 | +0.37 | −0.19 | 0 |
| O 3N site | B-MgO | −0.94 | 2.25 | 1.97 | 1.18 | −0.01 | −0.18 | 2 |
| | C-MgO | −1.18 | 1.98 | 1.87 | 1.16 | +0.49 | −0.13 | 1 |
| | N-MgO | −1.92 | 1.95 | 1.86 | 1.15 | +0.54 | −0.11 | 0 |

*refers to the distance between supported metal atom and the surface site (impurity site or O atom when metal is at O 3N site)

We have further analyzed the electronic structure of the systems with adsorbed CO. In the cases of both Pd (Fig. 5) and Au (Fig. 6) we observe that the *d*-states of the supported metal adatoms overlap with the $2\pi^*$ states of CO.[12] The presence of dopants significantly alters the states of supported adatoms and affects the strength of their interaction with CO. The alteration of the electronic states of the metal adatoms is especially prominent in the case of Au. Upon CO adsorption, the splitting of the d-states of Au supported by doped MgO(001) increases (the states of Au without CO are shown in Fig. 2). We have analyzed the charge distribution after CO adsorption and observed that for both Pd and Au there is a significant charge rearrangement. While after the adsorption CO bears a negative charge, the charge state of the adatoms depends on the type of dopant: negative for Pd and Au adsorbed at the B site, positive for all other impurities (Table 4). The impurities themselves are negatively charged. They bear slightly more charge in the presence of CO than without.

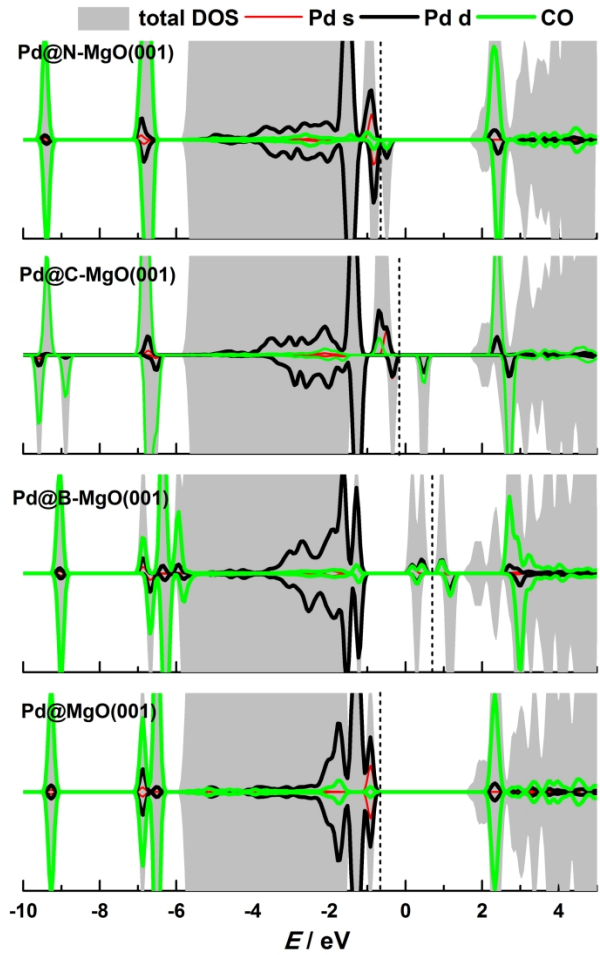

**Figure 5**. Projected densities of states of Pd supported by doped and pristine MgO(001) and CO molecule adsorbed on Pd atom. The vertical dashed lines indicate the highest occupied states.

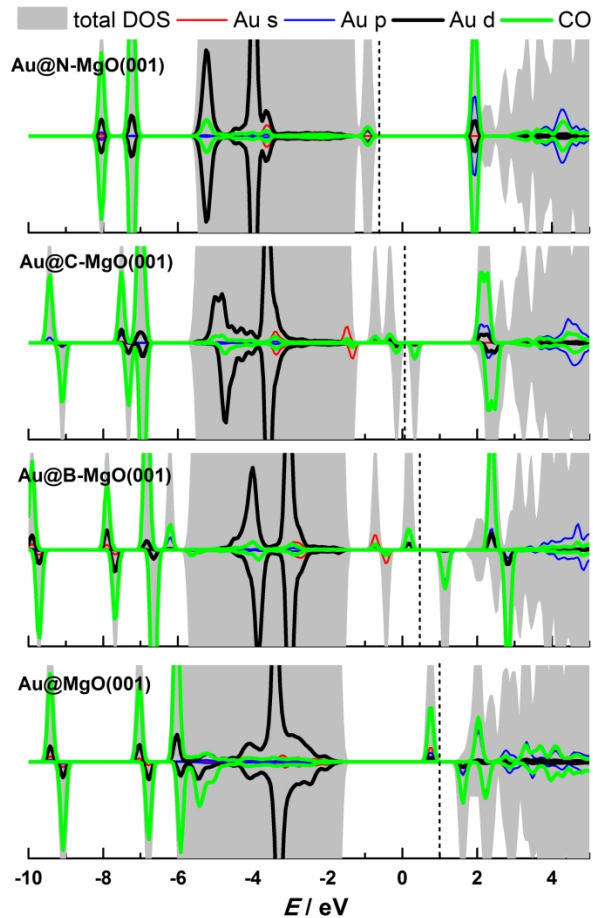

**Figure 6**. Projected densities of states of Au supported by doped and pristine MgO(001) and CO molecule adsorbed on Au atom. The vertical dashed lines indicate the highest occupied states.

We have previously shown that the alteration of the reactivity of doped MgO(001) is predominantly restricted to the impurity site.[42] This correlates with the calculated metal binding energies at oxygen sites of doped MgO(001) (O 2N and O 3N sites) and also the calculated charge transfer when compared to the values corresponding to adatom binding on pristine MgO(001) (Tables 1 and 3). Although we have shown that metal adatoms could easily get entrapped at the impurity site, we also analyzed the reactivity of Pd and Au adsorbed on doped MgO(001) away from the impurity site. In particular, we have studied the CO adsorption on Pd and Au located at the O 3N sites of doped MgO(001). The results for Pd show that such an adatom behaves just like Pd on pristine MgO(001). Moreover, the charge rearrangement upon the adsorption of CO is restricted to CO, Pd and the O 3N site, while the dopant has virtually no influence on the reactivity of the supported Pd atom (Table 4). The situation is drastically different in the case of Au, for which we observe a strong influence of dopants on the CO adsorption energies, increasing from B to N (Table 4). In the case of Au on B-doped MgO(001) the CO adsorption is similar to that on Au@MgO(001), but in the case of Au on N-doped MgO(001) the interaction becomes rather strong. Such drastic differences are also discernible from the projected densities of states (Fig. 7), showing a split of the *d*-states of Au on N-doped MgO(001) and a well localized d-band of Au on B-doped MgO(001), similar to that on pristine MgO(001) (Fig. 4).

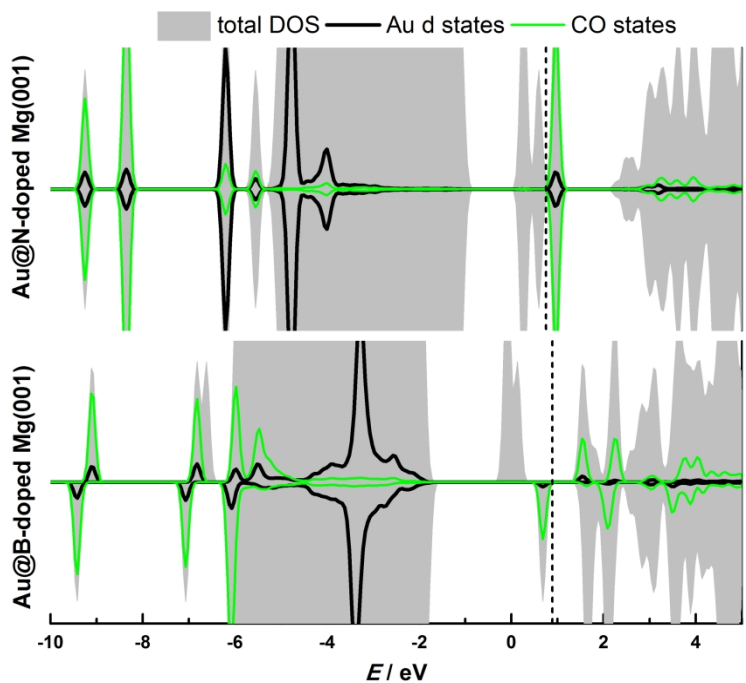

**Figure 7**. Projected densities of states of Au supported by B- and N-doped MgO(001) and CO molecule adsorbed on Au atom – the case of Au binding to O 3N site of doped MgO(001). Total densities of states are also included. The vertical dashed lines indicate the highest occupied states.

In parallel, we observe a significant charge rearrangement upon CO adsorption onto Au adatom at O 3N site, which is not present in the case of Pd adatom at the same adsorption site. In particular, the charge of Au on doped MgO(001) becomes more positive as going from B to N (Au bears a small negative charge in the case of B-doped MgO(001)), while the backdonation to the 2π* states of CO is reduced (Table 4). This is also in agreement with the decreasing C–O bond length when going from B- to N-doped surface. In addition, these changes are accompanied with charging of impurity atoms, which gets more prominent going from B to N. In order to clearly depict the difference in the behavior of Pd and Au, we show the charge redistribution upon the CO adsorption on adatoms located at different sites of N-doped MgO(001) (Fig. 8).

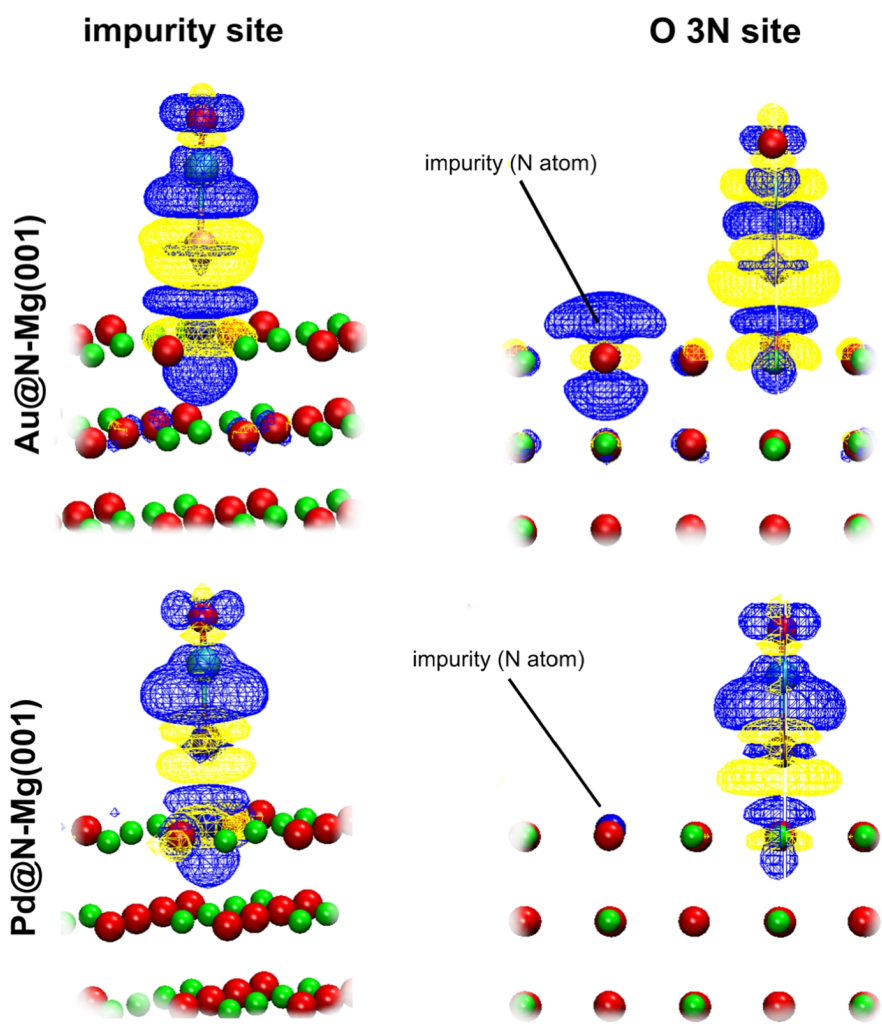

**Figure 8**. Charge difference plot for the case of CO adsorption on Au (top row) and Pd (bottom row) supported on N-doped MgO(001). Left column depicts the case of metal adsorbed on impurity site while right column is for the case metal on O 3N site. The isosurface values are ±0.0008 e Å$^{-3}$ (blue – build up of charge density, yellow – depletion of charge density).

The presented results clearly indicate that in the case of Au the impurity atom actively participates in the charge redistribution even when the adatom is relatively far away from it. This is not the case for Pd. Such distinct behaviors of Au and Pd are also seen in the case of adatom charging on thin oxide films supported by a metallic substrate, where charging occurs for Au but not Pd.[37] Furthermore, one can analyze the adsorption of CO considering a two-step process. First, one can deform the M@X-MgO(001) system to the configuration corresponding to the one with adsorbed CO. Then, the system can be frozen in such a configuration and the CO molecule can be attached to the adatom. We have modelled M@N-MgO(001) in the deformed configuration corresponding to the CO adsorption, on the metal atoms bound to N-site and O 3N site. In the case of Pd the deformation is very small (costs only ~0.02 eV) while the charge distribution is almost identical to that of the ground state configurations (Tables 2 and 3). The same holds for Au bound to the impurity site of N-MgO(001). Hence, in these cases the charge redistribution upon CO adsorption is governed by CO itself. However, in the case of Au bound to the O 3N site in the configuration corresponding to the CO adsorption, Au is much closer to the O site than in the case of pristine MgO(001) (Tables 1 and 4) and the deformation requires 0.56 eV. Shortening of the Au−O bond suggests a charge transfer from Au adatom and also the oxygen site. We found that in such a configuration the Au atom bears a positive charge and that the oxygen site losses some electrons, which are picked up by the impurity atom. This process reduces the magnetic moment of the system down to 0.23 $\mu_B$ (ground state magnetization is 2 $\mu_B$, Table 3). Once CO is adsorbed at Au, a further charge redistribution takes place and the ground state magnetization of the system goes to zero. However, we see that in this case the effect of CO on charge redistribution is indirect, acting *via* the deformation of the system, sufficient to induce the charge reorganization. However, the presence of an impurity, which can accept the charge, is required for the process to take place.

## 4. Conclusions

We report a significant alteration of reactivity of single Au and Pd adatoms on MgO(001) by surface doping with B, C and N. Impurity sites bind metal atoms much stronger than pristine MgO(001) terrace and can act as anchoring sites for Au and Pd adatoms. While the introduction of impurities can be a useful approach to obtain highly dispersed metal atoms with improved stability, their reactivity is significantly altered compared to metal atoms on pristine MgO(001). Pd supported by doped MgO(001) binds CO much weaker than Pd on pristine MgO(001). In contrast, gold supported by doped MgO(001) is much more reactive than gold adatom on MgO(001). In all the cases we see a charge backdonation to CO and a certain C–O elongation indicating the activation of the CO molecule. The change of the reactivity of the Pd adatom is practically seen only for the adatoms at impurity sites, while in the case of Au the reactivity of adatoms adsorbed far from impurity is also affected. We ascribe this effect to significant charge redistribution upon CO adsorption on Au adatoms, in which the impurity site actively participates. Such a charge rearrangement is not seen for Pd. The introduction of impurities into MgO(001) can be an elegant strategy to optimize the dispersion, stability and reactivity of supported metal atoms in the development of novel catalytic systems.

**Conflicts of interest**

There are no conflicts to declare.

**Acknowledgement**

I.A.P. acknowledges the support of the Serbian Ministry of Education, Science and Technological Development (Project No. III45014). N.V.S. acknowledges the support provided by Swedish Research Council through the project No. 2014-5993. The computations were performed on resources provided by the Swedish National Infrastructure for Computing (SNIC) at High Performance Computing Center North (HPC2N) at Umeå University. We also acknowledge the support from Carl Tryggers Foundation for Scientific Research.